# Search for $\eta_c(2S) \to 2(\pi^+\pi^-)$ and improved measurement of $\chi_{cJ} \to 2(\pi^+\pi^-)$

M. Ablikim *et al.*[*]
(BESIII Collaboration)

We search for the hadronic decay $\eta_c(2S) \to 2(\pi^+\pi^-)$ in the $\psi(3686) \to \gamma\eta_c(2S)$ radiative decay using $(27.12 \pm 0.14) \times 10^8$ $\psi(3686)$ events collected by the BESIII detector at the BEPCII collider. No significant signal is found, and the upper limit of $\mathcal{B}[\psi(3686) \to \gamma\eta_c(2S)]\mathcal{B}[\eta_c(2S) \to 2(\pi^+\pi^-)]$ is determined to be $1.43 \times 10^{-6}$ at the 90% confidence level. Using $\psi(3686) \to \gamma\chi_{cJ}$ transitions, we also measure the branching fractions of $\mathcal{B}[\chi_{cJ(J=0,1,2)} \to 2(\pi^+\pi^-)]$, which are $\mathcal{B}[\chi_{c0} \to 2(\pi^+\pi^-)] = (2.127 \pm 0.002(\text{stat}) \pm 0.101(\text{syst}))\%$, $\mathcal{B}[\chi_{c1} \to 2(\pi^+\pi^-)] = (0.685 \pm 0.001(\text{stat}) \pm 0.031(\text{syst}))\%$, and $\mathcal{B}[\chi_{c2} \to 2(\pi^+\pi^-)] = (1.153 \pm 0.001(\text{stat}) \pm 0.063(\text{syst}))\%$.

## I. INTRODUCTION

The $\eta_c(2S)$ state is the radial excited state of $\eta_c(1S)$ and was first observed by Belle in $B^\pm \to K^\pm\eta_c(2S)$ using the decay $\eta_c(2S) \to K^0_S K^\pm\pi^\mp$ [1]. Subsequently, this state was confirmed in the two-photon process $\gamma\gamma \to \eta_c(2S)$ by *BABAR* [2,3], CLEO [4], and Belle [5], and in the double-charmonium production process $e^+e^- \to J/\psi + c\bar{c}$ by Belle [6] and *BABAR* [7]. BESIII first observed the M1 transition process $\psi(3686) \to \gamma\eta_c(2S)$ using the $\eta_c(2S) \to K\bar{K}\pi$ decay mode [8]. Currently, there are only eight decay modes of the $\eta_c(2S)$ observed experimentally, with the uncertainties of all the measurements larger than 50%, and their summed decay width is around 5% of the total decay width of $\eta_c(2S)$ [9]. Therefore, searching for new decay modes is important for understanding the $\eta_c(2S)$ nature.

The ratio of the branching factions for $\psi(3686)$ and $J/\psi$ decaying to the same final states is predicted to be around 12% [10]. The $\eta_c(2S)$ and $\eta_c(1S)$ are spin-singlet partners of $\psi(3686)$ and $J/\psi$, and the ratio of $\frac{\mathcal{B}[\eta_c(2S)\to\text{hadrons}]}{\mathcal{B}[\eta_c(1S)\to\text{hadrons}]}$ is predicted to be 12% [11] or 100% [12]. Using experimental data on $\eta_c(2S)$ and $\eta_c(1S)$ decay to light hadron final states, authors of Ref. [13] found that the results of most decay modes differ from both of the two theoretical predictions, e.g., $\frac{\mathcal{B}[\eta_c(2S)\to K\bar{K}\pi]}{\mathcal{B}[\eta_c(1S)\to K\bar{K}\pi]} = 0.27^{+0.10}_{-0.07}$. New experimental results such as those on yet unexplored $\eta_c(2S) \to 2(\pi^+\pi^-)$ decays would give further insight into the $\eta_c(2S)$ decay dynamics.

The branching fractions of the hadronic decays $\chi_{cJ} \to 2(\pi^+\pi^-)$ ($J = 0$, 1, 2) were measured by the MARK I Collaboration in 1978 [14] and the BES Collaboration in 1999 [15], and their relative uncertainties are 7.7%, 34.2%, and 8.8%, respectively [9]. Improved measurements of these decay modes are needed for further understanding the $\chi_{cJ}$ decay dynamics.

In this analysis, we present a study of $\eta_c(2S) \to 2(\pi^+\pi^-)$ and $\chi_{cJ} \to 2(\pi^+\pi^-)$ in the radiative decays $\psi(3686) \to \gamma\eta_c(2S)/\chi_{cJ}$ and measure their branching fractions. Datasets collected by the BESIII detector at the BEPCII collider corresponding to $(27.12 \pm 0.14) \times 10^8$ $\psi(3686)$ events [16] produced at the center-of-mass (c.m.) energy of 3.686 GeV are used. The integrated luminosity of the dataset is 3877 $\text{pb}^{-1}$, which is measured using Bhabha process [16]. Additional datasets collected at 3.65 GeV and 3.682 GeV with integrated luminosities of 401 $\text{pb}^{-1}$ and 395 $\text{pb}^{-1}$ [16], respectively, are used to estimate the continuum background contribution.

## II. DECTECTOR AND MONTE CARLO SIMULATION

The BESIII detector [17] records symmetric $e^+e^-$ collisions provided by the BEPCII storage ring [18], which operates in the c.m. energy ($\sqrt{s}$) range from 2.0 to 4.95 GeV, with a peak luminosity of $1 \times 10^{33}\ \text{cm}^{-2}\,\text{s}^{-1}$ achieved at $\sqrt{s} = 3.773$ GeV. BESIII has collected large data samples in this energy region [19]. The cylindrical core of the BESIII detector covers 93% of the full solid angle and consists of a helium-based multilayer drift chamber (MDC), a plastic scintillator time-of-flight system (TOF), and a CsI(Tl) electromagnetic calorimeter (EMC), which are all enclosed in a superconducting solenoidal magnet providing a 1.0 T magnetic field [20]. The solenoid is supported by an octagonal flux-return yoke of steel plates interleaved with resistive plate counter based muon identification modules. The charged-particle momentum resolution at 1 GeV/$c$ is 0.5%, and resolution of the specific ionization energy loss ($\mathrm{d}E/\mathrm{d}x$) in the MDC is 6% for electrons from Bhabha scattering. The EMC measures photon energies with a resolution of 2.5% (5%) at 1 GeV in the barrel (end cap) region. The time resolution of the TOF in the barrel region is 68 ps, while that in the end cap region is 110 ps. The end cap TOF system was upgraded in 2015 using multi-gap resistive plate chamber technology, providing a time resolution of 60 ps [21–23], which benefits 83% of the data used in this analysis.

[*]Full author list given at the end of the article.

A GEANT4-based [24] Monte Carlo (MC) simulation program, which includes the description of the detector geometry and response, is used to produce large simulated event samples. These samples are used to optimize the event selection criteria, determine the detection efficiency, and estimate background contributions. The generator KKMC [25] is used to model the beam energy spread and the initial state radiation (ISR) effect. Exclusive MC samples of $\psi(3686) \to \gamma\eta_c(2S)$ and $\psi(3686) \to \gamma\chi_{cJ}$ are generated following the angular distribution of $(1 + \lambda\cos^2\theta)$, where $\theta$ is the polar angle of the radiative photon in the rest frame of $\psi(3686)$ and $\lambda$ is set to 1 for $\eta_c(2S)$ and to $1, -1/3, 1/13$ for $\chi_{cJ}$ ($J = 0$, 1, 2) [26], respectively. The $\eta_c(2S) \to 2(\pi^+\pi^-)$ and $\chi_{cJ} \to 2(\pi^+\pi^-)$ events are generated with a uniform distribution in phase space (PHSP). Additional exclusive background MC samples of $\psi(3686) \to (\gamma_{\mathrm{FSR}})2(\pi^+\pi^-)$, $\psi(3686) \to (\gamma_{\mathrm{FSR}})\rho^0\pi^+\pi^-$, and $\psi(3686) \to \pi^0 2(\pi^+\pi^-)$ are generated according to PHSP to estimate their contributions, where $\gamma_{\mathrm{FSR}}$ is a photon radiated by a final-state pion. Here final state radiation (FSR) from charged final state particles is incorporated using PHOTOS [27]. Generic MC samples including $\psi(3686)$ production and continuum processes are used to further analyze background contributions. The known decay modes are modeled with BESEVTGEN [28], where the known branching fractions are taken from the Particle Data Group (PDG) [9], and the unknown decay modes are generated by LUNDCHARM [29].

## III. SELECTION CRITERIA

We search for $\eta_c(2S)$ via the radiative decay $\psi(3686) \to \gamma\eta_c(2S)$ with $\eta_c(2S) \to 2(\pi^+\pi^-)$. Candidates must contain four charged tracks and at least one photon. Each charged track detected in the MDC must have its polar angle ($\theta$) within the active region ($|\cos\theta| < 0.93$), where $\theta$ is defined with respect to the $z$-axis, which is the symmetry axis of the MDC. Each charged track must originate from the interaction point (IP), which means that the distance of the closest approach to the IP of each track is required to be within 10 cm in the $z$ direction and within 1 cm in the plane perpendicular to the $z$-axis. The $\mathrm{d}E/\mathrm{d}x$ and TOF information are used for particle identification (PID), where a variable $\chi^2_{\mathrm{PID}}(h)$ is determined for each track for hypothesis $h$, and $h$ is a pion, kaon, or proton.

The photon candidates are selected from EMC showers. The deposited energy is required to be larger than 25 MeV in the barrel of EMC ($|\cos\theta| < 0.8$), or 50 MeV in the end-cap of EMC ($0.86 < |\cos\theta| < 0.92$). The angle between a good photon candidate and the nearest charged track in EMC is required to be larger than 10°. The timing of the shower is required to be within [0, 700] ns after the reconstructed event start time to suppress noise and energy deposits unrelated to the event.

Candidate events must have exactly four charged tracks with net charge zero and at least one candidate photon. For each event, we calculate the sum of $\chi^2_{\mathrm{PID}}(h, i)$ for the four tracks, where $i$ is the track number. Events with $\sum_{i=1}^4 \chi^2_{\mathrm{PID}}(\pi, i)$ less than any other assumption $[\sum_{i=1}^4 \chi^2_{\mathrm{PID}}(h_i, i)]$ will be retained, where each $h_i$ refers to an alternate hypothesis for track number $i$. A vertex fit, constraining the tracks to a common vertex, is performed on the four charged tracks, and events which do not pass the vertex fit are rejected. The total four-momentum of the photon candidate and four charged tracks is constrained to the initial $\psi(3686)$ using a kinematic fit (4C). If there is more than one photon candidate, the photon with the minimum $\chi^2$ from the 4C fit ($\chi^2_{4\mathrm{C}}$) is selected, and $\chi^2_{4\mathrm{C}}$ is required to be less than 40 to suppress background. This requirement is optimized by maximizing $S/\sqrt{S+B}$, where $S$ and $B$ are the expected number of $\eta_c(2S)$ signal and background events, respectively.

Background events from the $\psi(3686) \to \pi^+\pi^- J/\psi$ process are removed by requiring the recoil masses of all $\pi^+\pi^-$ pairs be outside the $J/\psi$ mass region ($M^{\mathrm{rec}}_{\pi^+\pi^-} < 3$ GeV/$c^2$ or $M^{\mathrm{rec}}_{\pi^+\pi^-} > 3.2$ GeV/$c^2$), which is referred to as the $J/\psi$ veto. Events from $\psi(3686) \to \eta J/\psi$ where $\eta$ decays to $\gamma\pi^+\pi^-$ are removed using a similar method, where the recoil masses of all $\gamma\pi^+\pi^-$ combinations are required to be outside the $J/\psi$ mass region, referred to as the $\eta J/\psi$ veto. In order to reject events where a photon converts to a $e^+e^-$ pair and both of the $e^+e^-$ are misidentified as pions along with other possible backgrounds, the angles between all combinations of $\pi^+$ and $\pi^-$ in the laboratory frame ($\theta_{\pi^+\pi^-}$) are required to be within $-0.999 < \cos\theta_{\pi^+\pi^-} < 0.988$, which is called the $\gamma$ conversion veto.

## IV. BACKGROUND ESTIMATION

Analysis of the generic MC sample of $\psi(3686)$ decays using TopoAna [30] indicates that the backgrounds mainly come from two sources: (1) $\psi(3686) \to \gamma_{\mathrm{non\ radiative}} 2(\pi^+\pi^-)$, where $\gamma_{\mathrm{non\ radiative}}$ represents a fake photon or an FSR photon, and this process may have a possible intermediate state $\rho^0$; and (2) $\psi(3686) \to \pi^0 2(\pi^+\pi^-)$ with $\pi^0 \to \gamma\gamma$. The remaining backgrounds distribute smoothly over the $2(\pi^+\pi^-)$ invariant mass spectrum.

### A. Background events from $\psi(3686) \to \gamma_{\text{non radiative}} 2(\pi^+\pi^-)$

Background from $\psi(3686) \to 2(\pi^+\pi^-)$ with a fake photon satisfying the 4C fit forms a $2(\pi^+\pi^-)$ peak shifted below the nominal $\psi(3686)$ mass [9], which makes it hard to separate this background from the $\eta_c(2S)$ signal. This shift does not occur when we perform a kinematic fit where the measured energy of the photon is not used (3C fit). In the 3C fit, the $\eta_c(2S)$ signal peak is similar with that in the 4C fit, while the $\psi(3686) \to 2(\pi^+\pi^-)$ background is more separated as shown in Fig. 1. Thus, the invariant mass of $2(\pi^+\pi^-)$ obtained from 3C fit ($M^{3C}_{2(\pi^+\pi^-)}$) is used to obtain the $\eta_c(2S)$ signal.

The background events of $\psi(3686) \to 2(\pi^+\pi^-)$ with an FSR photon have the same final states as the signal events. The FSR fraction, which is defined as $R_{\text{FSR}} = N_{\text{FSR}}/N_{\text{nonFSR}}$, may differ between data and MC simulation, and thus may need to be corrected using a FSR correction factor. Here, $N_{\text{FSR}}$ ($N_{\text{nonFSR}}$) is the number of events with (without) an FSR photon after the event selections. The FSR correction factor is studied using a control sample $\psi(3686) \to \gamma\chi_{c0}$, $\chi_{c0} \to (\gamma_{\text{FSR}})2(\pi^+\pi^-)$. The event selection criteria of the control sample are similar to those of the signal sample, except that at least two photon candidates are required. The softer photon is regarded as the FSR photon, and the energy of the FSR photon is allowed to vary in the 3C fit. The main background for the control sample is $\psi(3686) \to \pi^0 2(\pi^+\pi^-)$ since it has the same final state as the control sample. Events with $0.115\ \text{GeV}/c^2 < M_{\gamma\gamma} < 0.150\ \text{GeV}/c^2$ are rejected, where $M_{\gamma\gamma}$ is the invariant mass of the two photons. The energy of the photon with larger energy is required to be larger than 0.2 GeV to suppress background events from $\psi(3686) \to \gamma\chi_{c1,2}$, $\chi_{c1,2} \to (\gamma_{\text{FSR}})2(\pi^+\pi^-)$.

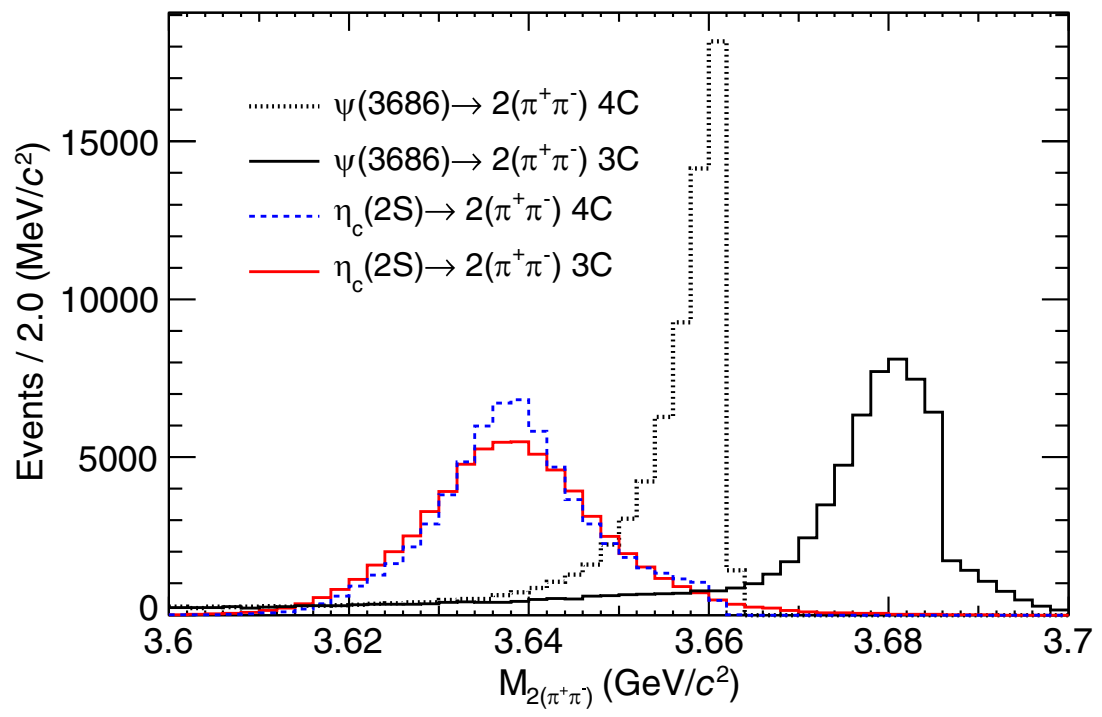

FIG. 1. The invariant mass distributions of $2(\pi^+\pi^-)$ from $\eta_c(2S) \to 2(\pi^+\pi^-)$ signal events using the 3C fit (red solid line) and the 4C fit (blue dashed line), and those from background events of $\psi(3686) \to 2(\pi^+\pi^-)$ using the 3C fit (black solid line) and the 4C fit (black dotted line).

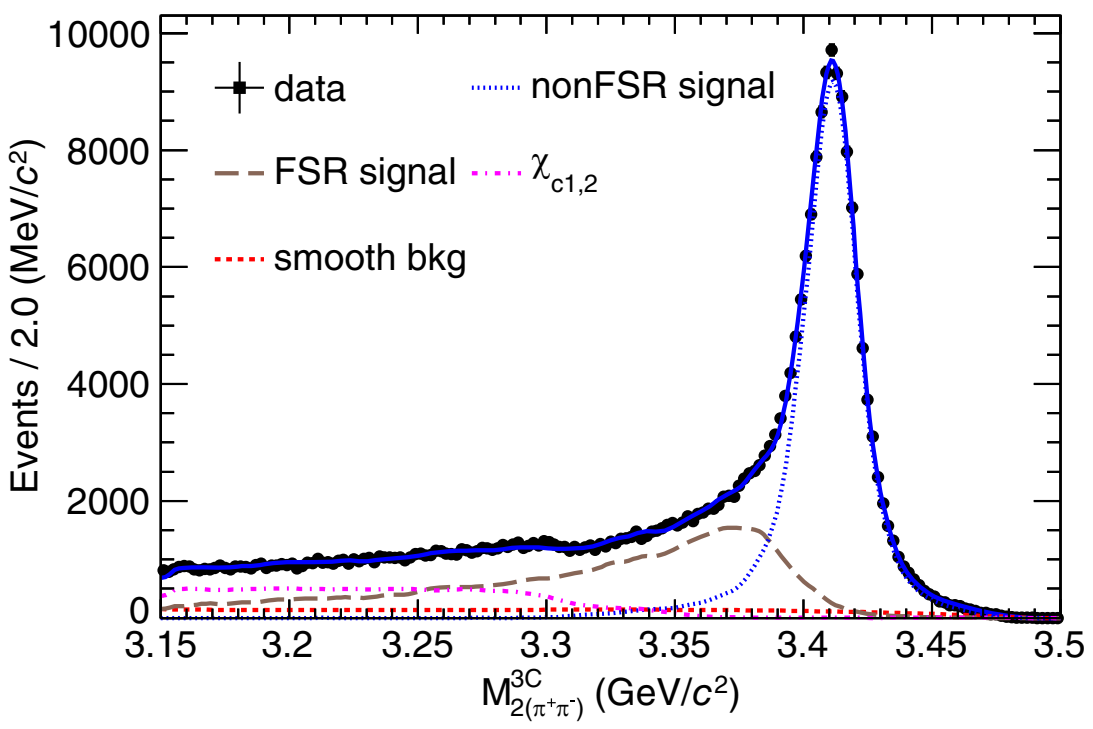

FIG. 2. Fit to the distribution of $M^{3C}_{2(\pi^+\pi^-)}$ from the selected $\psi(3686) \to \gamma\chi_{c0}$, $\chi_{c0} \to (\gamma_{\text{FSR}})2(\pi^+\pi^-)$ events. The black dots with error bars are data, the blue solid line is the total fit result, the blue dotted line represents the non-FSR component, the brown long-dashed line stands for the FSR component, the pink dash-dotted line represents the background from $\chi_{c1,2} \to 2(\pi^+\pi^-)$, and the red dashed line is the remaining smooth backgrounds.

The FSR fraction of MC ($R^{\text{MC}}_{\text{FSR}}$) is 0.323. The FSR fraction from data ($R^{\text{data}}_{\text{FSR}}$) is obtained by fitting the $M^{3C}_{2(\pi^+\pi^-)}$ distribution. The signals with and without FSR events are described by a shape from the $\psi(3686) \to \gamma\chi_{c0}$, $\chi_{c0} \to 2(\pi^+\pi^-)$ MC sample and convolved with a Gaussian function to account for the resolution difference between data and MC simulation. The parameters of the Gaussian function are floated in the fit. The background components $\psi(3686) \to \gamma\chi_{c1,2}$, $\chi_{c1,2} \to 2(\pi^+\pi^-)$ are determined from MC simulated events, and the ratio of its yield over the $\chi_{c0}$ component is fixed to the ratio obtained from MC simulation. The distribution of remaining background is modeled by a second order polynomial. The fit result is shown in Fig. 2. From the fit, we obtain $R^{\text{data}}_{\text{FSR}} = N^{\text{fit}}_{\text{FSR}}/N^{\text{fit}}_{\text{nonFSR}} = 0.647 \pm 0.006$, where $N^{\text{fit}}_{\text{FSR}}$ and $N^{\text{fit}}_{\text{nonFSR}}$ are the fitted yields with and without FSR events, respectively. Thus, the FSR correction factor is $f_{\text{FSR}} = R^{\text{data}}_{\text{FSR}}/R^{\text{MC}}_{\text{FSR}} = 2.00 \pm 0.02$, where the uncertainty is statistical.

The background shape of $\psi(3686) \to \gamma_{\text{non radiative}} 2(\pi^+\pi^-)$ is described by the sum of MC simulated shapes of $\psi(3686) \to 2(\pi^+\pi^-)$, $\rho^0\pi^+\pi^-$, $\gamma_{\text{FSR}}2(\pi^+\pi^-)$, and $\gamma_{\text{FSR}}\rho^0\pi^+\pi^-$, where events containing $\gamma_{\text{FSR}}$ are scaled by $f_{\text{FSR}}$, and the ratio of events with and without intermediate state $\rho^0$ is fixed [9].

### B. Background events from $\psi(3686) \to \pi^0 2(\pi^+\pi^-)$

The background from $\psi(3686) \to \pi^0 2(\pi^+\pi^-)$ is estimated from data. Events with $\gamma\gamma 2(\pi^+\pi^-)$ are selected using the selection criteria similar to those when selecting $\gamma 2(\pi^+\pi^-)$, except that two photons are required, and if there are more than two photons, the photon pair with the minimum $\chi^2$ from a 5C kinematic fit (4C fit for the photon pair and the four pions plus a $\pi^0$ mass constraint) is chosen. The $\chi^2$ from the 5C fit is required to be less than 50. The background distribution of $\psi(3686) \to \pi^0 2(\pi^+\pi^-)$ versus $M^{3C}_{2(\pi^+\pi^-)}$ is estimated using

$$\left(\frac{\mathrm{d}N}{\mathrm{d}M^{3C}_{2(\pi^+\pi^-)}}\right) = \left(\frac{\mathrm{d}N}{\mathrm{d}M^{5C}_{2(\pi^+\pi^-)}}\right) \times \frac{\varepsilon^{3C}_{\gamma 2(\pi^+\pi^-)}}{\varepsilon^{5C}_{\pi^0 2(\pi^+\pi^-)}}, \qquad (1)$$

where $(\frac{\mathrm{d}N}{\mathrm{d}M^{5C}_{2(\pi^+\pi^-)}})$ is the number of events in each $M^{5C}_{2(\pi^+\pi^-)}$ bin, $M^{5C}_{2(\pi^+\pi^-)}$ is the invariant mass of $2(\pi^+\pi^-)$ obtained from the 5C fit after passing the $\pi^0 2(\pi^+\pi^-)$ selection, $\varepsilon^{3C}_{\gamma 2(\pi^+\pi^-)}$ and $\varepsilon^{5C}_{\pi^0 2(\pi^+\pi^-)}$ are the $2(\pi^+\pi^-)$ efficiencies of the $\psi(3686) \to \pi^0 2(\pi^+\pi^-)$ MC simulated events passing the $\gamma 2(\pi^+\pi^-)$ and $\pi^0 2(\pi^+\pi^-)$ selections, respectively.

### C. Background events from continuum process

The background contribution from the continuum process $e^+e^- \to 2(\pi^+\pi^-)$ (including the FSR events) is obtained directly from MC simulation, and checked using datasets taken at c.m. energy ($E_{\mathrm{c.m.}}$) of 3.65 GeV. Events with FSR photons are corrected by $f_{\mathrm{FSR}}$ described in Sec. IV A.

## V. SIGNAL DETERMINATION

The signal yields are determined by a binned maximum likelihood fit to the $M^{3C}_{2(\pi^+\pi^-)}$ distribution, as shown in Fig. 3. The lineshapes of $\chi_{cJ}$ and $\eta_c(2S)$ are described by

$$[E^3_\gamma \times BW(m) \times f_d(E_\gamma)] \otimes G, \qquad (2)$$

where $m$ is the mass of $2(\pi^+\pi^-)$, $E_\gamma$ is the energy of the transition photon in the rest frame of $\psi(3686)$, $BW(m)$ is the Breit-Wigner function, $f_d(E_\gamma)$ is the function to damp the diverging tail from $E^3_\gamma$, and $G$ is a Gaussian resolution function describing the detector resolution. The $f_d(E_\gamma)$ by the KEDR Collaboration [31] is $E_0^2/[E_\gamma E_0 + (E_\gamma - E_0)^2]$, where $E_0 = [m^2_{\psi(3686)} - m^2_{\chi_{cJ}/\eta_c(2S)}]/[2m_{\psi(3686)}]$ is the peaking energy of the transition photon. For $\chi_{cJ}$, a double Gaussian function is used for the resolution function $G$, and its parameters are obtained directly from the fit, while for $\eta_c(2S)$, a Gaussian function is used, and its parameters are fixed to the values extrapolated from the $\chi_{cJ}$ results assuming a linear energy dependence. The shapes of background components are described above. For $\psi(3686) \to \pi^0 2(\pi^+\pi^-)$, the shape and the number of the events are fixed to the distribution obtained from data. For $\psi(3686) \to (\gamma_{\mathrm{FSR}}) 2(\pi^+\pi^-)$ and continuum backgrounds, the shapes are from MC simulations convolved with a Gaussian function whose parameters are floated, and the numbers of events are also free. The events with FSR photons in MC simulations are corrected using $f_{\mathrm{FSR}}$. The remaining background is smooth and is described by an ARGUS function [32] added by a piecewise polynomial whose parameters and number of events are free. The piecewise polynomial consists of a 2nd order polynomial in the range $[3.6, 3.67]$ GeV$/c^2$ and another 2nd order polynomial in $[3.67, 3.68]$ GeV$/c^2$, and they are required to be continuous at 3.67 GeV$/c^2$. A fit to the generic MC sample indicates that the input and output of the numbers of $\eta_c(2S)$ and $\chi_{cJ}$ signal events are statistically consistent.

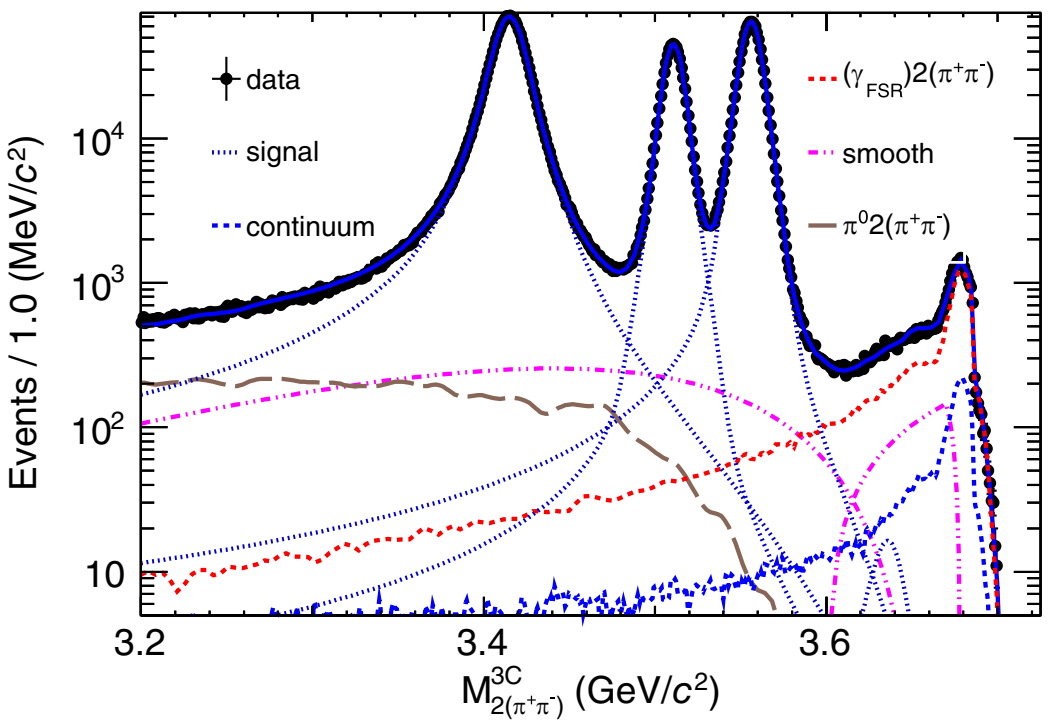

FIG. 3. The invariant mass distribution of $2(\pi^+\pi^-)$ after the 3C kinematic fit in the whole fit range. The black dots with error bars are data, the blue solid line is the total fit result, the blue dotted lines are the $\eta_c(2S)$ and $\chi_{cJ}$ signal shapes, the brown long-dashed line show the background from $\psi(3686) \to \pi^0 2(\pi^+\pi^-)$, the red dashed line is the background from $\psi(3686) \to (\gamma_{\mathrm{FSR}}) 2(\pi^+\pi^-)$, the blue dashed-dotted line shows the background from the continuum process, and the magenta dash-dot-doted lines represent two smooth background components (ARGUS function and piecewise polynomial).

The signal yields obtained from the fit are listed in Table I. The $\chi^2/\mathrm{ndf}$ value of the fit in the range $[3.6, 3.7]$ GeV$/c^2$ is 193.6/88, where ndf is the number of degrees of freedom. The statistical significance of $\eta_c(2S)$ signal is calculated to be $2.3\sigma$, using the difference of the logarithmic likelihoods [33], $-2\ln(\mathcal{L}_0/\mathcal{L}_{\max})$, where $\mathcal{L}_{\max}$ and $\mathcal{L}_0$ are the logarithmic likelihoods with and without the $\eta_c(2S)$ signal component, respectively.

The branching fractions of $\chi_{cJ} \to 2(\pi^+\pi^-)$ are calculated using

$$\mathcal{B}[\chi_{cJ} \to 2(\pi^+\pi^-)] = \frac{N^{\mathrm{sig}}_{\mathrm{data}}}{N^{\mathrm{tot}}_{\psi(3686)} \times \mathcal{B}(\psi(3686) \to \gamma\chi_{cJ}) \times \varepsilon(\chi_{cJ})}, \qquad (3)$$

where $N^{\mathrm{sig}}_{\mathrm{data}}$ is the fitted signal yield, $\mathcal{B}(\psi(3686) \to \gamma\chi_{cJ})$ is the branching fraction of $\psi(3686) \to \gamma\chi_{cJ}$ [9], and $\varepsilon(\chi_{cJ})$ is the signal detection efficiency determined by signal MC simulation. The fitted signal yields, the signal efficiencies, and the calculated branching fractions are listed in Table I. Since the significance of $\eta_c(2S)$ signal is less than $3\sigma$, the upper limit on the number of signal events ($N_{\rm U.L.}$) is determined at the 90% confidence level (CL) from

$$\int_0^{N_{\rm U.L.}} \mathcal{L}(x)\mathrm{d}x = 0.9 \int_0^{+\infty} \mathcal{L}(x)\mathrm{d}x, \tag{4}$$

where $x$ is the assumed signal yield and $\mathcal{L}(x)$ is the logarithmic likelihood of the data assuming $x$ signal events. The 90% CL upper limit of $\mathcal{B}[\psi(3686) \to \gamma\eta_c(2S)]\mathcal{B}[\eta_c(2S) \to 2(\pi^+\pi^-)]$ is calculated by $\frac{N_{\rm U.L.}}{N_{\psi(3686)}^{\rm tot} \times \varepsilon[\eta_c(2S)]}$, where $\varepsilon[\eta_c(2S)]$ is the signal detection efficiency of $\eta_c(2S)$. The upper limit of the number of signal events and the branching fraction are listed in Table I.

TABLE I. The signal yields ($N_{\rm data}^{\rm sig}$), the detection efficiencies ($\varepsilon$), and the calculated branching fractions ($\mathcal{B}$) for $\chi_{cJ} \to 2(\pi^+\pi^-)$, and the signal yield, the 90% CL upper limit of the signal yield ($N_{\rm U.L.}$), the detection efficiency, and the upper limit of the branching fraction ($\mathcal{B}_{\rm U.L.}$) for $\psi(3686) \to \gamma\eta_c(2S)$, $\eta_c(2S) \to 2(\pi^+\pi^-)$. For $\mathcal{B}$, the first uncertainties are statistical and the second uncertainties are systematic. For $N_{\rm U.L.}$ and $\mathcal{B}_{\rm U.L.}$, the systematic uncertainty is included.

| Channel | $N_{\rm data}^{\rm sig}$ | $\varepsilon$ (%) | $\mathcal{B}$ (%) |
|---|---|---|---|
| $\chi_{c0} \to 2(\pi^+\pi^-)$ | $(1946 \pm 2) \times 10^3$ | 34.3 | $2.127 \pm 0.002 \pm 0.101$ |
| $\chi_{c1} \to 2(\pi^+\pi^-)$ | $(670 \pm 1) \times 10^3$ | 36.8 | $0.685 \pm 0.001 \pm 0.031$ |
| $\chi_{c2} \to 2(\pi^+\pi^-)$ | $(1042 \pm 1) \times 10^3$ | 34.8 | $1.153 \pm 0.001 \pm 0.063$ |
| Channel | $N_{\rm data}^{\rm sig}$ [$N_{\rm U.L.}$] | $\varepsilon$ (%) | $\mathcal{B}$ [$\mathcal{B}_{\rm U.L.}$] ($\times 10^{-6}$) |
| $\psi(3686) \to \gamma\eta_c(2S), \eta_c(2S) \to 2(\pi^+\pi^-)$ | $461 \pm 188$ [1240] | 30.0 | $0.53 \pm 0.22 \pm 0.42$ [1.43] |

## VI. SYSTEMATIC UNCERTAINTIES

The systematic uncertainties of $\mathcal{B}[\chi_{cJ} \to 2(\pi^+\pi^-)]$ and $\mathcal{B}[\psi(3686) \to \gamma\eta_c(2S)]\mathcal{B}[\eta_c(2S) \to 2(\pi^+\pi^-)]$ are divided into two categories: multiplicative systematic uncertainties and additive systematic uncertainties, which are described below.

### A. Multiplicative systematic uncertainties

The systematic uncertainty of the tracking and PID efficiencies is estimated from the difference between data and MC simulation in the single-track reconstruction efficiency. The correction factor is defined as the weighted average $f_{\rm corr} = [\sum_i^{N_{\rm sel}} \prod_j^4 w_{ij}(p_t, \cos\theta)]/N_{\rm sel}$, where $N_{\rm sel}$ is the number of signal MC events that survive the signal selection, $i$ and $j$ run over the surviving events and the four charged tracks in each event, $w_{ij}$ is the charged track reconstruction weight factor in bins of $(p_t, \cos\theta)$, and $p_t$ is the transverse momentum of the track. The values of $w_{ij}$ as a function of $(p_t, \cos\theta)$ are obtained using a control sample of $J/\psi \to \pi^+\pi^-\pi^0$. The corrected efficiency is then calculated as $\varepsilon_{\rm corr} = \varepsilon \times f_{\rm corr}$. The difference between the corrected and nominal efficiencies is taken as the systematic uncertainty.

Based on the studies of photon detection efficiency using control samples $J/\psi \to \rho^0\pi^0$ and $e^+e^- \to \gamma\gamma$ [34], the systematic uncertainty due to photon reconstruction is assigned to be 1% per photon.

The decays $\chi_{cJ} \to 2(\pi^+\pi^-)$ and $\eta_c(2S) \to 2(\pi^+\pi^-)$ may have a possible intermediate state $\rho^0\pi^+\pi^-$ or $f_0(980)f_0(980)$ with $\rho^0/f_0(980) \to \pi^+\pi^-$, which are not taken into account in the PHSP MC samples. Thus, MC samples containing these intermediate states are generated to calculate the efficiencies, and the maximum differences between the new and the nominal efficiencies are taken as the systematic uncertainties.

In the kinematic fit, the helix parameters of charged tracks in MC samples have been corrected to improve the consistency between data and MC simulations [35]. The differences of efficiencies with and without the helix parameter correction are taken as the systematic uncertainties.

The uncertainty of number of $\psi(3686)$ events is 0.5% [16]. The systematic uncertainties from the branching fractions of $\psi(3686) \to \gamma\chi_{cJ}$ are taken from the PDG [9].

### B. Additive systematic uncertainties

There are three requirements used to veto the background events: the $J/\psi$ veto, $\eta J/\psi$ veto, and gamma conversion veto. The uncertainties from these requirements are estimated separately. For the $J/\psi$ veto, the lower bound of the requirement is varied from 3.0 GeV$/c^2$ to 2.95 or 3.05 GeV$/c^2$, or the upper bound from 3.2 GeV$/c^2$ to 3.15 or 3.25 GeV$/c^2$. For the $\eta J/\psi$ veto, the lower bound of the requirement is varied from 3.0 GeV$/c^2$ to 2.98 or 3.05 GeV$/c^2$. For the gamma conversion veto, the requirement is varied from 0.988 to 0.983 or 0.993. For each veto, the maximum difference of the branching fraction is taken as the corresponding uncertainty.

An alternative damping function used by the CLEO Collaboration [36], $f_d(E_\gamma) = \exp(-E_\gamma^2/8\beta^2)$, is used to estimate the systematic uncertainty due to the damping function form, where $\beta$ is a free parameter and is required to be the same for $\eta_c(2S)$ and $\chi_{cJ}$. The differences in the signal yields between the two damping functions are taken as the systematic uncertainties. The uncertainties from the mass resolution and mass shift are estimated by varying the mass resolution and mass shift by $\pm 1\sigma$, and the maximum differences are taken as the systematic uncertainties. The $\eta_c(2S)$ signal resolution shape is varied from a Gaussian to a double Gaussian function, where the parameters are extrapolated using $\chi_{cJ}$ signal resolution results. The difference is taken as the systematic uncertainty due to the $\eta_c(2S)$ signal resolution.

TABLE II. Relative systematic uncertainties in the measurements of branching fractions (in %).

| Source | $\mathcal{B}[\chi_{c0} \to 2(\pi^+\pi^-)]$ | $\mathcal{B}[\chi_{c1} \to 2(\pi^+\pi^-)]$ | $\mathcal{B}[\chi_{c2} \to 2(\pi^+\pi^-)]$ |
|---|---|---|---|
| Tracking efficiency and PID | 1.4 | 1.7 | 1.9 |
| Photon reconstruction | 1.0 | 1.0 | 1.0 |
| Intermediate state | 2.0 | 1.7 | 1.7 |
| Kinematic fit | 0.4 | 0.8 | 1.0 |
| Number of $\psi(3686)$ events | 0.5 | 0.5 | 0.5 |
| Branching fraction | 2.1 | 2.5 | 2.1 |
| $J/\psi$ veto | 1.9 | 1.5 | 1.5 |
| $\eta J/\psi$ veto | 0.9 | 0.7 | 0.6 |
| Gamma conversion veto | 2.4 | 1.8 | 1.9 |
| Damping function form | 0.5 | 0.4 | 1.6 |
| Mass resolution and shift | 0.7 | 0.1 | 0.2 |
| Size of FSR factor | 0.1 | 0.1 | 0.1 |
| Ratio of $\psi(3686) \to \rho^0\pi^+\pi^-$ events | 0.1 | 0.1 | 0.1 |
| Number of $\psi(3686) \to \pi^0 2(\pi^+\pi^-)$ events | 0.1 | 0.1 | 0.1 |
| Shape of smooth background | 0.4 | 0.5 | 0.2 |
| Bin width | 0.1 | 0.1 | 0.1 |
| Interference term | 0.2 | 0.1 | 3.0 |
| Total | 4.8 | 4.5 | 5.5 |

The systematic uncertainties related to the background contributions are from $\psi(3686) \to \gamma_{\rm FSR} 2(\pi^+\pi^-)$, $\psi(3686) \to \rho^0\pi^+\pi^-$, and $\psi(3686) \to \pi^0 2(\pi^+\pi^-)$. The FSR factor is varied by $\pm 1\sigma$ or changed to results from other channels ($1.62 \pm 0.13$, $1.62 \pm 0.07$, and $1.70 \pm 0.10$, respectively) [37–39], the ratio of the number of events of $\psi(3686) \to \rho^0\pi^+\pi^-$ over $\psi(3686) \to 2(\pi^+\pi^-)$ is varied by $\pm 1\sigma$, and the number of $\psi(3686) \to \pi^0 2(\pi^+\pi^-)$ events is varied by $\pm 1\sigma$. For each background component, the largest difference is taken as the systematic uncertainty. The shape of smooth background is changed from an ARGUS function [32] to a second order polynomial, and the difference is taken as the systematic uncertainty from the smooth background shape.

The bin width of the fit is changed from 1 MeV to 0.5 MeV, and the differences in the fitted signal yields compared to the nominal results are taken as the systematic uncertainties. There is possible interference between $\chi_{cJ}$ signal and continuum background. Each of the $\chi_{cJ}$ signals is allowed to interfere with the continuum background separately, and the maximum difference is taken as the systematic uncertainty.

For the $\chi_{cJ} \to 2(\pi^+\pi^-)$ channels, all of the systematic uncertainties are summarized in Table II. For the $\eta_c(2S) \to 2(\pi^+\pi^-)$ channel, since the statistical significance is less than $3\sigma$, the upper limit of $\mathcal{B}[\psi(3686) \to \gamma\eta_c(2S)]\mathcal{B}[\eta_c(2S) \to 2(\pi^+\pi^-)]$ at the 90% CL is determined, and the systematic uncertainties are taken into account in two steps. First, among the additive systematic uncertainties described above, the highest upper limit at the 90% CL is kept. Then, to take the multiplicative systematic uncertainties and the rest additive systematic uncertainties into account, the corresponding likelihood curve is convolved with a Gaussian function with a width parameter equal to the quadratic sum of all of the rest additive systematic uncertainties and the corresponding total multiplicative systematic uncertainty. All of the multiplicative systematic uncertainties are summarized in Table III. Assuming that all the multiplicative sources are independent, the total multiplicative systematic uncertainty is obtained by adding them in quadrature. The 90% CL upper

TABLE III. The multiplicative systematic uncertainties in the measurement of $\mathcal{B}[\psi(3686) \to \gamma\eta_c(2S)]\mathcal{B}[\eta_c(2S) \to 2(\pi^+\pi^-)]$ (in %).

| Source | $\mathcal{B}[\psi(3686) \to \gamma\eta_c(2S)]$ $\mathcal{B}[\eta_c(2S) \to 2(\pi^+\pi^-)]$ |
|---|---|
| Tracking efficiency and PID | 1.2 |
| Photon reconstruction | 1.0 |
| Intermediate state | 1.5 |
| Kinematic fit | 1.2 |
| Number of $\psi(3686)$ events | 0.6 |
| Total | 2.6 |

limit is then obtained by solving Eq. (4), and the result is listed in Table I.

## VII. SUMMARY

Using $(27.12 \pm 0.14) \times 10^8$ $\psi(3686)$ events collected by the BESIII detector at the BEPCII collider, a search for the hadronic decay $\eta_c(2S) \to 2(\pi^+\pi^-)$ is performed. No significant $\eta_c(2S)$ signal is found. The 90% CL upper limit of $\mathcal{B}[\psi(3686) \to \gamma\eta_c(2S)]\mathcal{B}[\eta_c(2S) \to 2(\pi^+\pi^-)]$ is determined to be $1.43 \times 10^{-6}$. The branching fractions of $\chi_{cJ} \to 2(\pi^+\pi^-)$ are summarized in Table I and are consistent with the previous results [9] but with improved precision. The relative uncertainty for the branching fraction of $\chi_{c1} \to 2(\pi^+\pi^-)$ is improved by a factor of 9. Using the PDG values of $\mathcal{B}[\psi(3686) \to \gamma\eta_c(2S)] = (7 \pm 5) \times 10^{-4}$ and $\mathcal{B}[\eta_c(1S) \to 2(\pi^+\pi^-)] = (8.7 \pm 1.1) \times 10^{-3}$ [9], the ratio $\mathcal{B}[\eta_c(2S) \to 2(\pi^+\pi^-)]/\mathcal{B}[\eta_c(1S) \to 2(\pi^+\pi^-)]$ is calculated to be less than 28.4% at the 90% CL, where the uncertainties of the branching fractions from the PDG are taken into account as multiplicative systematic uncertainty sources. This agrees with the results presented in Ref. [13], where the prediction of $\frac{\mathcal{B}[\eta_c(2S)\to \text{hadrons}]}{\mathcal{B}[\eta_c(1S)\to \text{hadrons}]} \approx 1$ is questioned. $\mathcal{B}[\eta_c(2S) \to 2(\pi^+\pi^-)]/\mathcal{B}[\eta_c(1S) \to 2(\pi^+\pi^-)]$ being 100% [12] can be ruled out.

## ACKNOWLEDGMENTS

The BESIII Collaboration thanks the staff of BEPCII and the IHEP computing center for their strong support. This work is supported in part by National Key R&D Program of China under Contracts No. 2020YFA0406300, No. 2020YFA0406400, No. 2023YFA1606000; National Natural Science Foundation of China (NSFC) under Contracts No. 11635010, No. 11735014, No. 11935015, No. 11935016, No. 11935018, No. 12025502, No. 12035009, No. 12035013, No. 12061131003, No. 12192260, No. 12192261, No. 12192262, No. 12192263, No. 12192264, No. 12192265, No. 12221005, No. 12225509, No. 12235017, No. 12361141819; the Chinese Academy of Sciences (CAS) Large-Scale Scientific Facility Program; the CAS Center for Excellence in Particle Physics (CCEPP); Joint Large-Scale Scientific Facility Funds of the NSFC and CAS under Contract No. U2032108 and No. U1832207; 100 Talents Program of CAS; The Institute of Nuclear and Particle Physics (INPAC) and Shanghai Key Laboratory for Particle Physics and Cosmology; German Research Foundation DFG under Contracts No. 455635585, No. FOR5327, No. GRK 2149; Istituto Nazionale di Fisica Nucleare, Italy; Ministry of Development of Turkey under Contract No. DPT2006K-120470; National Research Foundation of Korea under Contract No. NRF-2022R1A2C1092335; National Science and Technology fund of Mongolia; National Science Research and Innovation Fund (NSRF) via the Program Management Unit for Human Resources & Institutional Development, Research and Innovation of Thailand under Contract No. B16F640076; Polish National Science Centre under Contract No. 2019/35/O/ST2/02907; The Swedish Research Council; U.S. Department of Energy under Contract No. DE-FG02-05ER41374.

[1] S. K. Choi *et al.* (Belle Collaboration), Phys. Rev. Lett. **89**, 102001 (2002).
[2] B. Aubert *et al.* (*BABAR* Collaboration), Phys. Rev. Lett. **92**, 142002 (2004).
[3] B. Aubert *et al.* (*BABAR* Collaboration), Phys. Rev. D **84**, 012004 (2011).
[4] D. M. Asner *et al.* (CLEO Collaboration), Phys. Rev. Lett. **92**, 142001 (2004).
[5] H. Nakazawa *et al.* (Belle Collaboration), Nucl. Phys. B, Proc. Suppl. **184**, 220 (2008).
[6] K. Abe *et al.* (Belle Collaboration), Phys. Rev. Lett. **89**, 142001 (2002).
[7] B. Aubert *et al.* (*BABAR* Collaboration), Phys. Rev. D **72**, 031101 (2005).
[8] M. Ablikim *et al.* (BESIII Collaboration), Phys. Rev. Lett. **109**, 042003 (2012).
[9] R. L. Workman *et al.* (Particle Data Group), Prog. Theor. Exp. Phys. **2022**, 083C01 (2022) and 2023 update.
[10] T. Appelquist and H. D. Politzer, Phys. Rev. Lett. **34**, 43 (1975).
[11] M. Anselmino, M. Genovese, and E. Predazzi, Phys. Rev. D **44**, 1597 (1991).
[12] K. T. Chao, Y. F. Gu, and S. F. Tuan, Commun. Theor. Phys. **25**, 471 (1996).
[13] H. P. Wang and C. Z. Yuan, Chin. Phys. C **46**, 071001 (2022).
[14] W. Tanenbaum *et al.*, Phys. Rev. D **17**, 1731 (1978).
[15] J. Z. Bai *et al.* (BES Collaboration), Phys. Rev. D **60**, 072001 (1999).
[16] M. Ablikim *et al.* (BESIII Collaboration), Chin. Phys. C **42**, 023001 (2018). With the same approach as for $\psi'$ events taken in 2009, the preliminary number of $\psi'$ events taken in 2009, 2012 and 2021 is determined to be $27.12 \times 10^8$ with an uncertainty of 0.5%.
[17] M. Ablikim *et al.* (BESIII Collaboration), Nucl. Instrum. Methods Phys. Res., Sect. A **614**, 345 (2010).
[18] C. H. Yu *et al.*, *Proceedings of IPAC2016, Busan, Korea* (JACoW, Geneva, Switzerland, 2016).
[19] M. Ablikim *et al.* (BESIII Collaboration), Chin. Phys. C **44**, 040001 (2020).

[20] K. X. Huang, Z. J. Li, Z. Qian, J. Zhu, H. Y. Li, Y. M. Zhang, S. S. Sun, and Z. Y. You, Nucl. Sci. Tech. **33**, 142 (2022).
[21] X. Li *et al.*, Radiat. Detect. Technol. Methods **1**, 13 (2017).
[22] X. Y. Guo *et al.*, Radiat. Detect. Technol. Methods **1**, 15 (2017).
[23] P. Cao *et al.*, Nucl. Instrum. Methods Phys. Res., Sect. A **953**, 163053 (2020).
[24] S. Agostinelli *et al.* (Geant4 Collaboration), Nucl. Instrum. Methods Phys. Res., Sect. A **506**, 250 (2003).
[25] S. Jadach, B. F. L. Ward, and Z. Was, Comput. Phys. Commun. **130**, 260 (2000); Phys. Rev. D **63**, 113009 (2001).
[26] W. M. Tanenbaum *et al.*, Phys. Rev. D **17**, 1731 (1978).
[27] E. R. Was, Phys. Lett. B **303**, 163 (1993).
[28] D. J. Lange, Nucl. Instrum. Methods Phys. Res., Sect. A **462**, 152 (2001); R. G. Ping, Chin. Phys. C **32**, 599 (2008).
[29] J. C. Chen, G. S. Huang, X. R. Qi, D. H. Zhang, and Y. S. Zhu, Phys. Rev. D **62**, 034003 (2000); R. L. Yang, R. G. Ping, and H. Chen, Chin. Phys. Lett. **31**, 061301 (2014).
[30] X. Y. Zhou, S. X. Du, G. Li, and C. P. Shen, Comput. Phys. Commun. **258**, 107540 (2021).
[31] V. V. Anashin *et al.*, Int. J. Mod. Phys. Conf. Ser. **02**, 188 (2011).
[32] H. Albrecht *et al.* (ARGUS Collaboration), Phys. Lett. B **340**, 217 (1994).
[33] S. S. Wilks, Ann. Math. Stat. **9**, 60 (1938).
[34] M. Ablikim *et al.* (BESIII Collaboration), Phys. Rev. D **81**, 052005 (2010).
[35] M. Ablikim *et al.* (BESIII Collaboration), Phys. Rev. D **87**, 012002 (2013).
[36] R. E. Mitchell *et al.* (CLEO Collaboration), Phys. Rev. Lett. **102**, 011801 (2009).
[37] M. Ablikim *et al.* (BESIII Collaboration), Phys. Rev. D **106**, 032014 (2022).
[38] M. Ablikim *et al.* (BESIII Collaboration), Phys. Rev. D **107**, 052007 (2023).
[39] M. Ablikim *et al.* (BESIII Collaboration), Phys. Rev. D **84**, 091102 (2011).

M. Ablikim,[1] M. N. Achasov,[4,c] P. Adlarson,[76] O. Afedulidis,[3] X. C. Ai,[81] R. Aliberti,[35] A. Amoroso,[75a,75c] Q. An,[72,58,a] Y. Bai,[57] O. Bakina,[36] I. Balossino,[29a] Y. Ban,[46,h] H.-R. Bao,[64] V. Batozskaya,[1,44] K. Begzsuren,[32] N. Berger,[35] M. Berlowski,[44] M. Bertani,[28a] D. Bettoni,[29a] F. Bianchi,[75a,75c] E. Bianco,[75a,75c] A. Bortone,[75a,75c] I. Boyko,[36] R. A. Briere,[5] A. Brueggemann,[69] H. Cai,[77] X. Cai,[1,58] A. Calcaterra,[28a] G. F. Cao,[1,64] N. Cao,[1,64] S. A. Cetin,[62a] J. F. Chang,[1,58] G. R. Che,[43] G. Chelkov,[36,b] C. Chen,[43] C. H. Chen,[9] Chao Chen,[55] G. Chen,[1] H. S. Chen,[1,64] H. Y. Chen,[20] M. L. Chen,[1,58,64] S. J. Chen,[42] S. L. Chen,[45] S. M. Chen,[61] T. Chen,[1,64] X. R. Chen,[31,64] X. T. Chen,[1,64] Y. B. Chen,[1,58] Y. Q. Chen,[34] Z. J. Chen,[25,i] Z. Y. Chen,[1,64] S. K. Choi,[10] G. Cibinetto,[29a] F. Cossio,[75c] J. J. Cui,[50] H. L. Dai,[1,58] J. P. Dai,[79] A. Dbeyssi,[18] R. E. de Boer,[3] D. Dedovich,[36] C. Q. Deng,[73] Z. Y. Deng,[1] A. Denig,[35] I. Denysenko,[36] M. Destefanis,[75a,75c] F. De Mori,[75a,75c] B. Ding,[67,1] X. X. Ding,[46,h] Y. Ding,[40] Y. Ding,[34] J. Dong,[1,58] L. Y. Dong,[1,64] M. Y. Dong,[1,58,64] X. Dong,[77] M. C. Du,[1] S. X. Du,[81] Y. Y. Duan,[55] Z. H. Duan,[42] P. Egorov,[36,b] Y. H. Fan,[45] J. Fang,[1,58] J. Fang,[59] S. S. Fang,[1,64] W. X. Fang,[1] Y. Fang,[1] Y. Q. Fang,[1,58] R. Farinelli,[29a] L. Fava,[75b,75c] F. Feldbauer,[3] G. Felici,[28a] C. Q. Feng,[72,58] J. H. Feng,[59] Y. T. Feng,[72,58] M. Fritsch,[3] C. D. Fu,[1] J. L. Fu,[64] Y. W. Fu,[1,64] H. Gao,[64] X. B. Gao,[41] Y. N. Gao,[46,h] Yang Gao,[72,58] S. Garbolino,[75c] I. Garzia,[29a,29b] L. Ge,[81] P. T. Ge,[19] Z. W. Ge,[42] C. Geng,[59] E. M. Gersabeck,[68] A. Gilman,[70] K. Goetzen,[13] L. Gong,[40] W. X. Gong,[1,58] W. Gradl,[35] S. Gramigna,[29a,29b] M. Greco,[75a,75c] M. H. Gu,[1,58] Y. T. Gu,[15] C. Y. Guan,[1,64] A. Q. Guo,[31,64] L. B. Guo,[41] M. J. Guo,[50] R. P. Guo,[49] Y. P. Guo,[12,g] A. Guskov,[36,b] J. Gutierrez,[27] K. L. Han,[64] T. T. Han,[1] F. Hanisch,[3] X. Q. Hao,[19] F. A. Harris,[66] K. K. He,[55] K. L. He,[1,64] F. H. Heinsius,[3] C. H. Heinz,[35] Y. K. Heng,[1,58,64] C. Herold,[60] T. Holtmann,[3] P. C. Hong,[34] G. Y. Hou,[1,64] X. T. Hou,[1,64] Y. R. Hou,[64] Z. L. Hou,[1] B. Y. Hu,[59] H. M. Hu,[1,64] J. F. Hu,[56,j] S. L. Hu,[12,g] T. Hu,[1,58,64] Y. Hu,[1] G. S. Huang,[72,58] K. X. Huang,[59] L. Q. Huang,[31,64] X. T. Huang,[50] Y. P. Huang,[1] Y. S. Huang,[59] T. Hussain,[74] F. Hölzken,[3] N. Hüsken,[35] N. in der Wiesche,[69] J. Jackson,[27] S. Janchiv,[32] J. H. Jeong,[10] Q. Ji,[1] Q. P. Ji,[19] W. Ji,[1,64] X. B. Ji,[1,64] X. L. Ji,[1,58] Y. Y. Ji,[50] X. Q. Jia,[50] Z. K. Jia,[72,58] D. Jiang,[1,64] H. B. Jiang,[77] P. C. Jiang,[46,h] S. S. Jiang,[39] T. J. Jiang,[16] X. S. Jiang,[1,58,64] Y. Jiang,[64] J. B. Jiao,[50] J. K. Jiao,[34] Z. Jiao,[23] S. Jin,[42] Y. Jin,[67] M. Q. Jing,[1,64] X. M. Jing,[64] T. Johansson,[76] S. Kabana,[33] N. Kalantar-Nayestanaki,[65] X. L. Kang,[9] X. S. Kang,[40] M. Kavatsyuk,[65] B. C. Ke,[81] V. Khachatryan,[27] A. Khoukaz,[69] R. Kiuchi,[1] O. B. Kolcu,[62a] B. Kopf,[3] M. Kuessner,[3] X. Kui,[1,64] N. Kumar,[26] A. Kupsc,[44,76] W. Kühn,[37] J. J. Lane,[68] L. Lavezzi,[75a,75c] T. T. Lei,[72,58] Z. H. Lei,[72,58] M. Lellmann,[35] T. Lenz,[35] C. Li,[47] C. Li,[43] C. H. Li,[39] Cheng Li,[72,58] D. M. Li,[81] F. Li,[1,58] G. Li,[1] H. B. Li,[1,64] H. J. Li,[19] H. N. Li,[56,j] Hui Li,[43] J. R. Li,[61] J. S. Li,[59] K. Li,[1] L. J. Li,[1,64] L. K. Li,[1] Lei Li,[48] M. H. Li,[43] P. R. Li,[38,k,l] Q. M. Li,[1,64] Q. X. Li,[50] R. Li,[17,31] S. X. Li,[12] T. Li,[50] W. D. Li,[1,64] W. G. Li,[1,a] X. Li,[1,64] X. H. Li,[72,58] X. L. Li,[50] X. Y. Li,[1,64] X. Z. Li,[59] Y. G. Li,[46,h] Z. J. Li,[59] Z. Y. Li,[79] C. Liang,[42] H. Liang,[72,58] H. Liang,[1,64] Y. F. Liang,[54] Y. T. Liang,[31,64] G. R. Liao,[14] Y. P. Liao,[1,64] J. Libby,[26] A. Limphirat,[60] C. C. Lin,[55] D. X. Lin,[31,64] T. Lin,[1] B. J. Liu,[1] B. X. Liu,[77] C. Liu,[34] C. X. Liu,[1] F. Liu,[1] F. H. Liu,[53] Feng Liu,[6] G. M. Liu,[56,j] H. Liu,[38,k,l] H. B. Liu,[15] H. H. Liu,[1] H. M. Liu,[1,64] Huihui Liu,[21] J. B. Liu,[72,58] J. Y. Liu,[1,64] K. Liu,[38,k,l] K. Y. Liu,[40] Ke Liu,[22]

L. Liu,[72,58] L. C. Liu,[43] Lu Liu,[43] M. H. Liu,[12,g] P. L. Liu,[1] Q. Liu,[64] S. B. Liu,[72,58] T. Liu,[12,g] W. K. Liu,[43] W. M. Liu,[72,58] X. Liu,[39] X. Liu,[38,k,l] Y. Liu,[81] Y. Liu,[38,k,l] Y. B. Liu,[43] Z. A. Liu,[1,58,64] Z. D. Liu,[9] Z. Q. Liu,[50] X. C. Lou,[1,58,64] F. X. Lu,[59] H. J. Lu,[23] J. G. Lu,[1,58] X. L. Lu,[1] Y. Lu,[7] Y. P. Lu,[1,58] Z. H. Lu,[1,64] C. L. Luo,[41] J. R. Luo,[59] M. X. Luo,[80] T. Luo,[12,g] X. L. Luo,[1,58] X. R. Lyu,[64] Y. F. Lyu,[43] F. C. Ma,[40] H. Ma,[79] H. L. Ma,[1] J. L. Ma,[1,64] L. L. Ma,[50] L. R. Ma,[67] M. M. Ma,[1,64] Q. M. Ma,[1] R. Q. Ma,[1,64] T. Ma,[72,58] X. T. Ma,[1,64] X. Y. Ma,[1,58] Y. Ma,[46,h] Y. M. Ma,[31] F. E. Maas,[18] M. Maggiora,[75a,75c] S. Malde,[70] Y. J. Mao,[46,h] Z. P. Mao,[1] S. Marcello,[75a,75c] Z. X. Meng,[67] J. G. Messchendorp,[13,65] G. Mezzadri,[29a] H. Miao,[1,64] T. J. Min,[42] R. E. Mitchell,[27] X. H. Mo,[1,58,64] B. Moses,[27] N. Yu. Muchnoi,[4,c] J. Muskalla,[35] Y. Nefedov,[36] F. Nerling,[18,e] L. S. Nie,[20] I. B. Nikolaev,[4,c] Z. Ning,[1,58] S. Nisar,[11,m] Q. L. Niu,[38,k,l] W. D. Niu,[55] Y. Niu,[50] S. L. Olsen,[64] Q. Ouyang,[1,58,64] S. Pacetti,[28b,28c] X. Pan,[55] Y. Pan,[57] A. Pathak,[34] Y. P. Pei,[72,58] M. Pelizaeus,[3] H. P. Peng,[72,58] Y. Y. Peng,[38,k,l] K. Peters,[13,e] J. L. Ping,[41] R. G. Ping,[1,64] S. Plura,[35] V. Prasad,[33] F. Z. Qi,[1] H. Qi,[72,58] H. R. Qi,[61] M. Qi,[42] T. Y. Qi,[12,g] S. Qian,[1,58] W. B. Qian,[64] C. F. Qiao,[64] X. K. Qiao,[81] J. J. Qin,[73] L. Q. Qin,[14] L. Y. Qin,[72,58] X. P. Qin,[12,g] X. S. Qin,[50] Z. H. Qin,[1,58] J. F. Qiu,[1] Z. H. Qu,[73] C. F. Redmer,[35] K. J. Ren,[39] A. Rivetti,[75c] M. Rolo,[75c] G. Rong,[1,64] Ch. Rosner,[18] S. N. Ruan,[43] N. Salone,[44] A. Sarantsev,[36,d] Y. Schelhaas,[35] K. Schoenning,[76] M. Scodeggio,[29a] K. Y. Shan,[12,g] W. Shan,[24] X. Y. Shan,[72,58] Z. J. Shang,[38,k,l] J. F. Shangguan,[16] L. G. Shao,[1,64] M. Shao,[72,58] C. P. Shen,[12,g] H. F. Shen,[1,8] W. H. Shen,[64] X. Y. Shen,[1,64] B. A. Shi,[64] H. Shi,[72,58] H. C. Shi,[72,58] J. L. Shi,[12,g] J. Y. Shi,[1] Q. Q. Shi,[55] S. Y. Shi,[73] X. Shi,[1,58] J. J. Song,[19] T. Z. Song,[59] W. M. Song,[34,1] Y. J. Song,[12,g] Y. X. Song,[46,h,n] S. Sosio,[75a,75c] S. Spataro,[75a,75c] F. Stieler,[35] Y. J. Su,[64] G. B. Sun,[77] G. X. Sun,[1] H. Sun,[64] H. K. Sun,[1] J. F. Sun,[19] K. Sun,[61] L. Sun,[77] S. S. Sun,[1,64] T. Sun,[51,f] W. Y. Sun,[34] Y. Sun,[9] Y. J. Sun,[72,58] Y. Z. Sun,[1] Z. Q. Sun,[1,64] Z. T. Sun,[50] C. J. Tang,[54] G. Y. Tang,[1] J. Tang,[59] M. Tang,[72,58] Y. A. Tang,[77] L. Y. Tao,[73] Q. T. Tao,[25,i] M. Tat,[70] J. X. Teng,[72,58] V. Thoren,[76] W. H. Tian,[59] Y. Tian,[31,64] Z. F. Tian,[77] I. Uman,[62b] Y. Wan,[55] S. J. Wang,[50] B. Wang,[1] B. L. Wang,[64] Bo Wang,[72,58] D. Y. Wang,[46,h] F. Wang,[73] H. J. Wang,[38,k,l] J. J. Wang,[77] J. P. Wang,[50] K. Wang,[1,58] L. L. Wang,[1] M. Wang,[50] N. Y. Wang,[64] S. Wang,[12,g] S. Wang,[38,k,l] T. Wang,[12,g] T. J. Wang,[43] W. Wang,[73] W. Wang,[59] W. P. Wang,[35,72,o] W. P. Wang,[72,58] X. Wang,[46,h] X. F. Wang,[38,k,l] X. J. Wang,[39] X. L. Wang,[12,g] X. N. Wang,[1] Y. Wang,[61] Y. D. Wang,[45] Y. F. Wang,[1,58,64] Y. L. Wang,[19] Y. N. Wang,[45] Y. Q. Wang,[1] Yaqian Wang,[17] Yi Wang,[61] Z. Wang,[1,58] Z. L. Wang,[73] Z. Y. Wang,[1,64] Ziyi Wang,[64] D. H. Wei,[14] F. Weidner,[69] S. P. Wen,[1] Y. R. Wen,[39] U. Wiedner,[3] G. Wilkinson,[70] M. Wolke,[76] L. Wollenberg,[3] C. Wu,[39] J. F. Wu,[1,8] L. H. Wu,[1] L. J. Wu,[1,64] X. Wu,[12,g] X. H. Wu,[34] Y. Wu,[72,58] Y. H. Wu,[55] Y. J. Wu,[31] Z. Wu,[1,58] L. Xia,[72,58] X. M. Xian,[39] B. H. Xiang,[1,64] T. Xiang,[46,h] D. Xiao,[38,k,l] G. Y. Xiao,[42] S. Y. Xiao,[1] Y. L. Xiao,[12,g] Z. J. Xiao,[41] C. Xie,[42] X. H. Xie,[46,h] Y. Xie,[50] Y. G. Xie,[1,58] Y. H. Xie,[6] Z. P. Xie,[72,58] T. Y. Xing,[1,64] C. F. Xu,[1,64] C. J. Xu,[59] G. F. Xu,[1] H. Y. Xu,[67,2,p] M. Xu,[72,58] Q. J. Xu,[16] Q. N. Xu,[30] W. Xu,[1] W. L. Xu,[67] X. P. Xu,[55] Y. C. Xu,[78] Z. S. Xu,[64] F. Yan,[12,g] L. Yan,[12,g] W. B. Yan,[72,58] W. C. Yan,[81] X. Q. Yan,[1,64] H. J. Yang,[51,f] H. L. Yang,[34] H. X. Yang,[1] T. Yang,[1] Y. Yang,[12,g] Y. F. Yang,[1,64] Y. F. Yang,[43] Y. X. Yang,[1,64] Z. W. Yang,[38,k,l] Z. P. Yao,[50] M. Ye,[1,58] M. H. Ye,[8] J. H. Yin,[1] Junhao Yin,[43] Z. Y. You,[59] B. X. Yu,[1,58,64] C. X. Yu,[43] G. Yu,[1,64] J. S. Yu,[25,i] T. Yu,[73] X. D. Yu,[46,h] Y. C. Yu,[81] C. Z. Yuan,[1,64] J. Yuan,[45] J. Yuan,[34] L. Yuan,[2] S. C. Yuan,[1,64] Y. Yuan,[1,64] Z. Y. Yuan,[59] C. X. Yue,[39] A. A. Zafar,[74] F. R. Zeng,[50] S. H. Zeng,[63a,63b,63c,63d] X. Zeng,[12,g] Y. Zeng,[25,i] Y. J. Zeng,[59] Y. J. Zeng,[1,64] X. Y. Zhai,[34] Y. C. Zhai,[50] Y. H. Zhan,[59] A. Q. Zhang,[1,64] B. L. Zhang,[1,64] B. X. Zhang,[1] D. H. Zhang,[43] G. Y. Zhang,[19] H. Zhang,[81] H. Zhang,[72,58] H. C. Zhang,[1,58,64] H. H. Zhang,[59] H. H. Zhang,[34] H. Q. Zhang,[1,58,64] H. R. Zhang,[72,58] H. Y. Zhang,[1,58] J. Zhang,[81] J. Zhang,[59] J. J. Zhang,[52] J. L. Zhang,[20] J. Q. Zhang,[41] J. S. Zhang,[12,g] J. W. Zhang,[1,58,64] J. X. Zhang,[38,k,l] J. Y. Zhang,[1] J. Z. Zhang,[1,64] Jianyu Zhang,[64] L. M. Zhang,[61] Lei Zhang,[42] P. Zhang,[1,64] Q. Y. Zhang,[34] R. Y. Zhang,[38,k,l] S. H. Zhang,[1,64] Shulei Zhang,[25,i] X. D. Zhang,[45] X. M. Zhang,[1] X. Y. Zhang,[50] Y. Zhang,[73] Y. Zhang,[1] Y. T. Zhang,[81] Y. H. Zhang,[1,58] Y. M. Zhang,[39] Yan Zhang,[72,58] Z. D. Zhang,[1] Z. H. Zhang,[1] Z. L. Zhang,[34] Z. Y. Zhang,[43] Z. Y. Zhang,[77] Z. Z. Zhang,[45] G. Zhao,[1] J. Y. Zhao,[1,64] J. Z. Zhao,[1,58] L. Zhao,[1] Lei Zhao,[72,58] M. G. Zhao,[43] N. Zhao,[79] R. P. Zhao,[64] S. J. Zhao,[81] Y. B. Zhao,[1,58] Y. X. Zhao,[31,64] Z. G. Zhao,[72,58] A. Zhemchugov,[36,b] B. Zheng,[73] B. M. Zheng,[34] J. P. Zheng,[1,58] W. J. Zheng,[1,64] Y. H. Zheng,[64] B. Zhong,[41] X. Zhong,[59] H. Zhou,[50] J. Y. Zhou,[34] L. P. Zhou,[1,64] S. Zhou,[6] X. Zhou,[77] X. K. Zhou,[6] X. R. Zhou,[72,58] X. Y. Zhou,[39] Y. Z. Zhou,[12,g] A. N. Zhu,[64] J. Zhu,[43] K. Zhu,[1] K. J. Zhu,[1,58,64] K. S. Zhu,[12,g] L. Zhu,[34] L. X. Zhu,[64] S. H. Zhu,[71] T. J. Zhu,[12,g] W. D. Zhu,[41] Y. C. Zhu,[72,58] Z. A. Zhu,[1,64] J. H. Zou,[1] and J. Zu[72,58]

(BESIII Collaboration)

[1]*Institute of High Energy Physics, Beijing 100049, People's Republic of China*
[2]*Beihang University, Beijing 100191, People's Republic of China*
[3]*Bochum Ruhr-University, D-44780 Bochum, Germany*

[4]Budker Institute of Nuclear Physics SB RAS (BINP), Novosibirsk 630090, Russia
[5]Carnegie Mellon University, Pittsburgh, Pennsylvania 15213, USA
[6]Central China Normal University, Wuhan 430079, People's Republic of China
[7]Central South University, Changsha 410083, People's Republic of China
[8]China Center of Advanced Science and Technology, Beijing 100190, People's Republic of China
[9]China University of Geosciences, Wuhan 430074, People's Republic of China
[10]Chung-Ang University, Seoul, 06974, Republic of Korea
[11]COMSATS University Islamabad, Lahore Campus, Defence Road, Off Raiwind Road, 54000 Lahore, Pakistan
[12]Fudan University, Shanghai 200433, People's Republic of China
[13]GSI Helmholtzcentre for Heavy Ion Research GmbH, D-64291 Darmstadt, Germany
[14]Guangxi Normal University, Guilin 541004, People's Republic of China
[15]Guangxi University, Nanning 530004, People's Republic of China
[16]Hangzhou Normal University, Hangzhou 310036, People's Republic of China
[17]Hebei University, Baoding 071002, People's Republic of China
[18]Helmholtz Institute Mainz, Staudinger Weg 18, D-55099 Mainz, Germany
[19]Henan Normal University, Xinxiang 453007, People's Republic of China
[20]Henan University, Kaifeng 475004, People's Republic of China
[21]Henan University of Science and Technology, Luoyang 471003, People's Republic of China
[22]Henan University of Technology, Zhengzhou 450001, People's Republic of China
[23]Huangshan College, Huangshan 245000, People's Republic of China
[24]Hunan Normal University, Changsha 410081, People's Republic of China
[25]Hunan University, Changsha 410082, People's Republic of China
[26]Indian Institute of Technology Madras, Chennai 600036, India
[27]Indiana University, Bloomington, Indiana 47405, USA
[28a]INFN Laboratori Nazionali di Frascati, I-00044, Frascati, Italy
[28b]INFN Sezione di Perugia, I-06100, Perugia, Italy
[28c]University of Perugia, I-06100, Perugia, Italy
[29a]INFN Sezione di Ferrara, I-44122, Ferrara, Italy
[29b]University of Ferrara, I-44122, Ferrara, Italy
[30]Inner Mongolia University, Hohhot 010021, People's Republic of China
[31]Institute of Modern Physics, Lanzhou 730000, People's Republic of China
[32]Institute of Physics and Technology, Peace Avenue 54B, Ulaanbaatar 13330, Mongolia
[33]Instituto de Alta Investigación, Universidad de Tarapacá, Casilla 7D, Arica 1000000, Chile
[34]Jilin University, Changchun 130012, People's Republic of China
[35]Johannes Gutenberg University of Mainz, Johann-Joachim-Becher-Weg 45, D-55099 Mainz, Germany
[36]Joint Institute for Nuclear Research, 141980 Dubna, Moscow region, Russia
[37]Justus-Liebig-Universitaet Giessen, II. Physikalisches Institut, Heinrich-Buff-Ring 16, D-35392 Giessen, Germany
[38]Lanzhou University, Lanzhou 730000, People's Republic of China
[39]Liaoning Normal University, Dalian 116029, People's Republic of China
[40]Liaoning University, Shenyang 110036, People's Republic of China
[41]Nanjing Normal University, Nanjing 210023, People's Republic of China
[42]Nanjing University, Nanjing 210093, People's Republic of China
[43]Nankai University, Tianjin 300071, People's Republic of China
[44]National Centre for Nuclear Research, Warsaw 02-093, Poland
[45]North China Electric Power University, Beijing 102206, People's Republic of China
[46]Peking University, Beijing 100871, People's Republic of China
[47]Qufu Normal University, Qufu 273165, People's Republic of China
[48]Renmin University of China, Beijing 100872, People's Republic of China
[49]Shandong Normal University, Jinan 250014, People's Republic of China
[50]Shandong University, Jinan 250100, People's Republic of China
[51]Shanghai Jiao Tong University, Shanghai 200240, People's Republic of China
[52]Shanxi Normal University, Linfen 041004, People's Republic of China
[53]Shanxi University, Taiyuan 030006, People's Republic of China
[54]Sichuan University, Chengdu 610064, People's Republic of China
[55]Soochow University, Suzhou 215006, People's Republic of China
[56]South China Normal University, Guangzhou 510006, People's Republic of China
[57]Southeast University, Nanjing 211100, People's Republic of China

[58]*State Key Laboratory of Particle Detection and Electronics, Beijing 100049, Hefei 230026, People's Republic of China*
[59]*Sun Yat-Sen University, Guangzhou 510275, People's Republic of China*
[60]*Suranaree University of Technology, University Avenue 111, Nakhon Ratchasima 30000, Thailand*
[61]*Tsinghua University, Beijing 100084, People's Republic of China*
[62a]*Turkish Accelerator Center Particle Factory Group, Istinye University, 34010, Istanbul, Turkey*
[62b]*Near East University, Nicosia, North Cyprus, 99138, Mersin 10, Turkey*
[63a]*University of Bristol, H H Wills Physics Laboratory*
[63b]*Tyndall Avenue*
[63c]*Bristol*
[63d]*BS8 1TL*
[64]*University of Chinese Academy of Sciences, Beijing 100049, People's Republic of China*
[65]*University of Groningen, NL-9747 AA Groningen, The Netherlands*
[66]*University of Hawaii, Honolulu, Hawaii 96822, USA*
[67]*University of Jinan, Jinan 250022, People's Republic of China*
[68]*University of Manchester, Oxford Road, Manchester, M13 9PL, United Kingdom*
[69]*University of Muenster, Wilhelm-Klemm-Strasse 9, 48149 Muenster, Germany*
[70]*University of Oxford, Keble Road, Oxford OX13RH, United Kingdom*
[71]*University of Science and Technology Liaoning, Anshan 114051, People's Republic of China*
[72]*University of Science and Technology of China, Hefei 230026, People's Republic of China*
[73]*University of South China, Hengyang 421001, People's Republic of China*
[74]*University of the Punjab, Lahore-54590, Pakistan*
[75a]*University of Turin and INFN, I-10125, Turin, Italy*
[75b]*University of Eastern Piedmont, I-15121, Alessandria, Italy*
[75c]*INFN, I-10125, Turin, Italy*
[76]*Uppsala University, Box 516, SE-75120 Uppsala, Sweden*
[77]*Wuhan University, Wuhan 430072, People's Republic of China*
[78]*Yantai University, Yantai 264005, People's Republic of China*
[79]*Yunnan University, Kunming 650500, People's Republic of China*
[80]*Zhejiang University, Hangzhou 310027, People's Republic of China*
[81]*Zhengzhou University, Zhengzhou 450001, People's Republic of China*

[a]Deceased.
[b]Also at the Moscow Institute of Physics and Technology, Moscow 141700, Russia.
[c]Also at the Novosibirsk State University, Novosibirsk, 630090, Russia.
[d]Also at the NRC "Kurchatov Institute", PNPI, 188300, Gatchina, Russia.
[e]Also at Goethe University Frankfurt, 60323 Frankfurt am Main, Germany.
[f]Also at Key Laboratory for Particle Physics, Astrophysics and Cosmology, Ministry of Education; Shanghai Key Laboratory for Particle Physics and Cosmology; Institute of Nuclear and Particle Physics, Shanghai 200240, People's Republic of China.
[g]Also at Key Laboratory of Nuclear Physics and Ion-beam Application (MOE) and Institute of Modern Physics, Fudan University, Shanghai 200443, People's Republic of China.
[h]Also at State Key Laboratory of Nuclear Physics and Technology, Peking University, Beijing 100871, People's Republic of China.
[i]Also at School of Physics and Electronics, Hunan University, Changsha 410082, China.
[j]Also at Guangdong Provincial Key Laboratory of Nuclear Science, Institute of Quantum Matter, South China Normal University, Guangzhou 510006, China.
[k]Also at MOE Frontiers Science Center for Rare Isotopes, Lanzhou University, Lanzhou 730000, People's Republic of China.
[l]Also at Lanzhou Center for Theoretical Physics, Lanzhou University, Lanzhou 730000, People's Republic of China.
[m]Also at the Department of Mathematical Sciences, IBA, Karachi 75270, Pakistan.
[n]Also at Ecole Polytechnique Federale de Lausanne (EPFL), CH-1015 Lausanne, Switzerland.
[o]Also at Helmholtz Institute Mainz, Staudinger Weg 18, D-55099 Mainz, Germany.
[p]Also at School of Physics, Beihang University, Beijing 100191, China.